# Influencing Factors of the FLASH Effect: Unveiling the Importance of Free Radicals

**Short Running Title:** The importance of Free Radicals in FLASH effect

Yan Zhang[1]†, Chenyang Huang[1]†, Ankang Hu[2,3], Yucheng Wang[1], Wanyi Zhou[2,3], Jiaqi Qiu[4], Jian Wang [4], Qibin Fu[1]*, Tuchen Huang[1]*, Hao Zha[2], Wei Wang[1], Xiaowu Deng [5], Junli Li[2,3]

1 Sino-French Institute of Nuclear Engineering and Technology, Sun Yat-sen University, Zhuhai, 519082, China
2 Department of Engineering Physics, Tsinghua University, Beijing, China
3 Key Laboratory of Particle & Radiation Imaging, Tsinghua University, Ministry of Education, Beijing, China
4 THACCEV COMPANY LIMITED, China
5 Department of Radiation Oncology, Sun Yat-sen University Cancer Center, State Key Laboratory of Oncology in South China, Collaborative Innovation Center for Cancer Medicine, Guangzhou, China

†These authors contributed equally to this work.

*Correspondence:
Qibin Fu: fuqibin@mail.sysu.edu.cn
Tuchen Huang: huangtuchen@mail.sysu.edu.cn;

## ABSTRACT

***Purpose:*** The precise physical conditions that elicit the FLASH effect remain inconclusive. In this work, we investigated the biological impacts of varying irradiation parameters utilizing a murine model of acute intestinal toxicity. Our aim was to elucidate the critical factors responsible for inducing the FLASH effect, focusing on the role of free radicals through a combination of simulation and experimental approaches.
***Methods and Materials:*** The whole abdomen of C57BL/6 mice was irradiated with 6 MeV electron beam. The endpoint was acute intestinal toxicity quantified by histological score. Total doses ranging from 6 to 15 Gy were evaluated. The impact of the mean dose rate (MDR) was assessed in the range of 40 to 900 Gy/s. Dose per pulse (DPP) of 0.5 Gy and 3 Gy were compared at a total dose of 15 Gy and a fixed MDR. The concentration of protein hydroperoxides (ROOH) and recombination of peroxyl radicals (ROO●) were simulated. Further comparisons were conducted by incorporating the antioxidant amifostine.
***Results:*** When varying total doses with a constant MDR of 900 Gy/s, the FLASH effect was not observed until the dose reached 15 Gy. The histological score of FLASH group at the dose of 15 Gy was close to that of CONV group at the dose of 12 Gy. For a total dose of 15 Gy and varying MDR, the FLASH effect was observed only when MDR reached 100 Gy/s, corresponding to dose delivery time of 150 ms. The magnitude of the FLASH effect reached saturation after the MDR exceeded 200 Gy/s (i.e., dose delivery time less than 75 ms). For a dose of 15 Gy and an MDR of 150 Gy/s, no significant difference in biological effect was observed between low DPP (0.5 Gy/pulse) and high DPP (3 Gy/pulse). The simulation results indicated that the fraction of ROO● recombination remained nearly zero at conventional dose rates (CONV), irrespective of the dose administered. In contrast, for FLASH irradiation, the recombination fraction increased almost linearly with the dose. The ROOH concentration of FLASH at 15 Gy is very close to that of CONV at 12 Gy, which is consistent with the experimental result. For FLASH condition, the relationship between the ROO● recombination fraction and dose delivery time exhibited an S-shaped curve. Notably, the dose delivery time corresponding to 50% change in the recombination fraction was approximately 300 ms for a dose of 15 Gy, which is consistent with the results of varying MDR experiment. The fraction of ROO● recombination approaches the maximum when dose delivery time is less than 100 ms, very close to the value of 75 ms when the magnitude of the FLAHS effect reached maximum. The addition of amifostine effectively eliminated the difference between FLASH group and CONV group.
***Conclusions:*** Our research indicates that the critical requirement for observing the sparing effect with a specified significance at the biological endpoint is the administration of an adequate dose within the time window of the radical reaction, irrespective of the pulse structure. The relationship between free radical recombination and exposure time and dose were quantitatively simulated, which were in good agreement with the experimental results. Additionally, the important role of free radical was verified after introducing antioxidants, suggesting that the generation and recombination of free radicals are pivotal factors influencing the FLASH sparing effect.

## Introduction

FLASH radiotherapy (FLASH-RT) is distinguished by its ultra-high dose rate delivery (≥ 40 Gy/s), which is several orders of magnitude greater than that of conventional radiotherapy (CONV-RT). FLASH-RT has been shown to maintain tumor control comparable to CONV-RT while mitigating radiation-induced damage to normal tissues, a phenomenon known as the FLASH effect. This effect has been documented across a range of tissues and biological endpoints using various types of radiation(1–12). Nonetheless, there are also studies reporting the absence of the FLASH effect or even contrary outcomes. Research indicates that the FLASH effect is associated with various physical parameters, including mean dose rate, total dose, pulse dose rate, pulse dose, and irradiation volume(9, 13). We performed a systematic review of published experimental studies to elucidate the relationship between the FLASH effect, radiation dose, and mean dose rate(13). Regrettably, numerous studies lack comprehensive details regarding irradiation beam parameters, thereby constraining the systematic comparison and analysis of various factors. Consequently, it is crucial to delineate the conditions that induce the FLASH effect and to elucidate the relationship between different physical variables. Furthermore, the robustness and reproducibility of experiments are essential for advancing the exploration of the underlying mechanisms of the FLASH effect.

Typically, the irradiation duration for FLASH-RT is less than one second. The radiochemical phase following ionization and excitation significantly influences the resultant biological effects, encompassing both the heterogeneous (at the time scale of nanoseconds to microseconds) and homogeneous (at the time scale of microseconds to milliseconds) phases of free radicals(14–17). This aspect is especially critical for low linear energy transfer (LET) radiation, such as X-rays and electrons, which primarily induce damage through indirect mechanisms mediated by free radicals(18). For the same radiation dose, ultra-high dose-rate irradiation generates a transiently elevated concentration of free radicals over a brief duration, in contrast to conventional dose-rate irradiation. This phenomenon can modify the chemical pathways responsible for cell death and tissue damage, aligning with the radical recombination hypothesis proposed by researchers in recent years(19). The FLASH technique is anticipated to produce a high instantaneous concentration of peroxyl radicals, thereby increasing the likelihood of radical recombination. Free radical recombination reactions yield non-radical products, consequently mitigating the detrimental effects associated with free radicals. Subsequently, the free radical recombination-antioxidant hypothesis was proposed(20), suggesting that antioxidants can compete with peroxyl free radicals, thereby elucidating the differential responses observed between tumor and normal tissues.

To investigate these phenomena, we examined the impact of mean dose rate, total dose, pulse dose, and pulse dose rate on acute intestinal toxicity in mice using electron beam irradiation. Additionally, we explored the role of free radicals in the FLASH effect by incorporating antioxidants into the experimental framework. Our research indicates that the critical requirement for observing the FLASH effect with a specified significance at the biological endpoint is the administration of an adequate dose within the temporal framework of the radical reaction, irrespective of the pulse structure. The interplay between free radical recombination and both dose and irradiation duration were further investigated through simulation. And no significant difference in intestinal damage was found between FLASH-RT and CONV-RT following the introduction of antioxidants, suggesting that the generation and recombination of free radicals are pivotal factors influencing the FLASH effect.

## Materials and Methods

## Mice handling

Female C57BL/6 mice, aged six to eight weeks, were procured from the Laboratory Animal Resources Center at Tsinghua University (THU-LARC) for experimental purposes. The mice were maintained on a diet of standard chow and had unrestricted access to acidified water. Housing conditions included cages with a regulated 12-hour light/dark cycle. All procedures involving the mice received approval from the Institutional Animal Care and Use Committee (IACUC) at THU-LARC (Form ID: THU-LARC-2024-021).

## Irradiation setup

Irradiation was conducted utilizing an electron linear accelerator from the Department of Engineering Physics at Tsinghua University, which provides a vertically irradiated electron beam with an energy of 6 MeV. Whole abdomen irradiation was administered to the mice by positioning a collimator at the accelerator's exit window, thereby achieving a field size of $45 \times 45\ mm^2$ (refer to Fig. 1A and 1B). The radiation dose for each mouse was quantified using calibrated EBT3 films, with doses ranging from 6 to 15 Gy. During irradiation, mice were positioned on a board with a film placed both above and below the abdomen. The films were scanned 24 hours post-irradiation, and the validation of homogeneous dose coverage is illustrated in Fig. 1. Fig. 1E and 1F demonstrated that within the irradiation field, the doses administered in the vertical, longitudinal, and lateral directions achieved the planned levels of 8 Gy for CONV-RT and 16 Gy for FLASH-RT. The dose distribution was observed to be uniform. All conventional irradiation procedures were conducted at a dose rate of 0.05 Gy/s. The specific beam parameters for FLASH-RT are detailed in Table. 1.

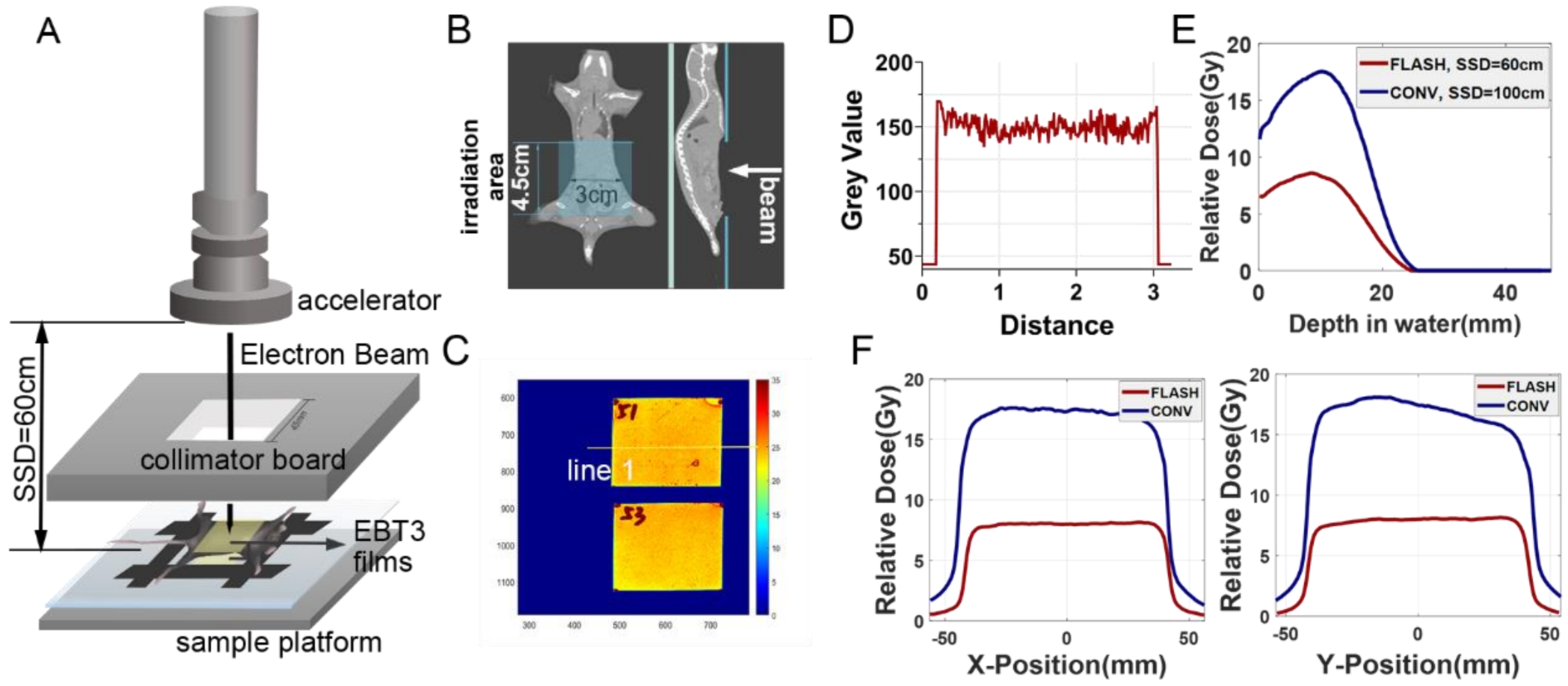

Fig. 1. The experimental platform employed for whole abdominal irradiation (WAI). (A) A schematic representation of the experimental setup for FLASH-RT. (B) Top and side computed tomography (CT) images of mice illustrating the irradiation status. (C) The spatial dose distribution was measured using an EBT3 film. (D) A straight line (line 1) was drawn on the film in Fig 1C to assess grey values at different points. The vertical dose along the central axis of the irradiation plane (E) and the relative dose variations in the longitudinal and lateral directions within the plane (F) were evaluated. The planned doses were 8 Gy for CONV-RT and 16 Gy for FLASH-RT.

Table. 1 The beam parameters of FLASH-RT

| | Dose (Gy) | Mean dose rate (Gy/s) | Pulse frequency (Hz) | Pulse dose rate (Gy/s) | No. of pulses | Pulse dose (Gy) |
|---|---|---|---|---|---|---|
| Fig 2B | 6/9/12/15 | 900 | 300 | $1.00*10^6$ | 2/3/4/5 | 3 |
| Fig 2C | 15 | 40/100/200/750 | 13.33/33.33/66.67/250 | $1.00*10^6$ | 5 | 3 |
| Fig 2D | 15 | 150Gy/s | 300 | $2.00*10^5$ | 30 | 0.5 |
| | 15 | 150Gy/s | 50 | $1.00*10^6$ | 5 | 3 |

## Drug Treatment

Amifostine (MCE, HY-B0639) was prepared in sterile saline at 40 mg/mL and given intraperitoneally at 200 mg/kg. A prior study determined the median lethal dose (LD50) of Amifostine to be 833±20 mg/kg, a value significantly exceeding the dosage employed in our experiment (21). Consequently, the intrinsic toxicity of the drug can be considered minimal. Amifostine was administered 15-30 minutes prior to irradiation.

## Tissue Processing and Histological Analysis

A five-centimeter segment of the jejunum was excised, rinsed in phosphate-buffered saline (PBS), and subsequently fixed overnight in 4% paraformaldehyde (PFA). The following day, the intestinal segment was longitudinally incised along the mesenteric line using scissors. Subsequently, the edge of the jejunum was carefully wrapped around a toothpick using forceps, and the toothpick was gently rolled with fingers to coil the colon around it, forming a Swiss roll. The tissues fixed in PFA were then processed for paraffin embedding. Sections of 5 μm thickness were cut using a microtome and subsequently stained with hematoxylin and eosin(H&E).

For histological analysis, a semi-quantitative histological score was determined based on a modified Geobes score and Chiu's score (22, 23), evaluating six parameters, including submucosal inflammation, crypt structure, and crypt regeneration, inflammation cell infiltration in lamina propria, thickening of the muscularis mucosae and epithelial damage. Severity is graded on a scale from 0 (least severe) to 4 (most severe), as detailed in Table S1.

## Simulation

A simplified two-compartment model was used to study the chemical reactions and damage caused by FLASH irradiation, simulating interactions within the lipid and other molecular compartments(20). Reactants were categorized based on reaction types and rate constants. This model highlights the recombination of peroxyl radicals and their interactions with antioxidants across both compartments. The proportion of recombined peroxyl radicals was calculated to investigate the significance of recombination. The detailed computational procedures employed in this study were based on a previous study (20), which included the concentration of ROOH and the rates of radical recombination. The dose modification factor (DMF) is a metric utilized to quantify the efficacy of a substance in altering the biological effects of radiation. It is defined as the ratio of doses from different radiation types that produce equivalent biological effects.

## Statistical analysis

Data were presented as mean ± SD. One-way analysis of variance (ANOVA) was used to assess statistical significance among multiple groups, with a significance level set at $p < 0.05$. All analyses were performed in GraphPad Prism 8.0.2.

## Results

### FLASH radiotherapy mitigates radiation-induced intestinal injury by administering an adequate dose within a proper time window.

To examine the influence of physical factors on the sparing effects of FLASH-RT, it is essential to maintain consistency in the biological variables of the mice, including strain, sex, and age. The intestinal response to acute radiation-induced injury is most pronounced between three to four days post-exposure. Consequently, acute intestinal toxicity was assessed 72 hours following irradiation using histological analysis. Due to potential variability in endpoints, this study employed a comprehensive histological scoring system previously described in the literature(22, 23). As illustrated in Fig. 2A, as the radiation dose increased from 6 to 15 Gy, the severity of intestinal injury correspondingly escalated. Specifically, intestinal injury exhibited phenomenon such as inflammatory cell infiltration, the intact extent of crypt structure, crypt regenerative capacity, etc. (Table S1). To ascertain the critical role of dose in observing the FLASH sparing effect, the MDR was set to a constant value of 900 Gy/s. No biological differences between FLASH-RT and CONV-RT were observed until the dose increased to 15 Gy (Fig. 2B), at which the damage of FLASH-RT was significantly lower than that of CONV-RT (Fig. 2B). This indicates that a sufficient dose is necessary to observe the sparing effect.

While the radiation dose is sufficient to induce substantial damage, the delivery time must be short enough to activate the FLASH sparing effect. To determine the time window, the effects were evaluated with MDR ranging from 40 Gy/s to 750 Gy/s. As shown in Fig. 2C, a significant reduction in histological scores began to be observed when MDR increased to 100 Gy/s (Fig. 2C), corresponding to a delivery time of 150 ms. Furthermore, the magnitude of the FLASH sparing effect was found to increase with higher MDR, but reaching a maximum when MDR exceeds 200 Gy/s (Fig. 2C).

Additionally, we explored the impact of varying DPP on the observed FLASH sparing effect. At a total dose of 15 Gy and MDR of 150 Gy/s, the high-DPP irradiation was delivered with 5 pulses (3 Gy/pulse) while the low-DPP irradiation was delivered with 30 pulses (0.5 Gy/pulse). The FLASH sparing effect was observed in both cases and there was no significant difference in the magnitude of the sparing effect (Fig. 2D). The results indicate that the effect of DPP on FLASH sparing effect is less pronounced than that of dose and MDR.

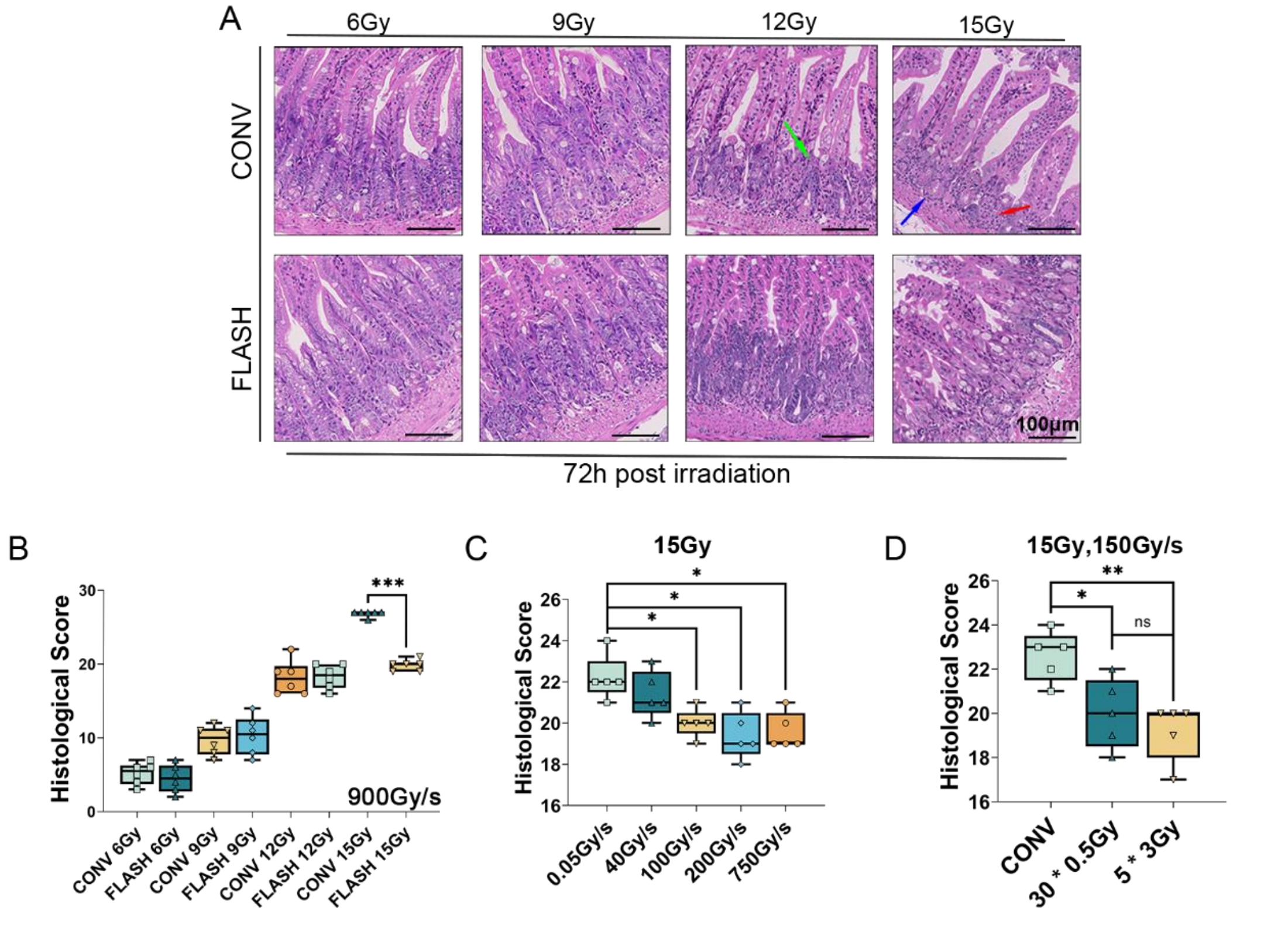

Fig. 2. The physical factors of radiation required for the FLASH sparing effect were observed at the biological endpoint. (A) Representative images of H&E-stained sections of jejunum segments at 72h after different doses of WAI. Different colors of arrows indicate different types of tissue injury mentioned previously, red: infiltration of inflammatory cells; blue: the architectural damage; green: a severe loss of numbers of crypts. Scale Bar= 100 μm (B) Histological score of intestine tissues varied between FLASH-RT (900Gy/s) and CONV-RT across dose escalation (n=6 per group). (C) Changes of intestinal histological scores with different mean dose rates after irradiation (n=5 per group). (D) Histological scores of intestinal tissues after irradiation at different pulse dose rates (n=5 per group), the specific beam parameters are shown in Table. 1. *p<0.05, **p<0.01, ***p<0.001.

## Simulation of free radical production and recombination

To explore the underlying mechanisms of the sparing effect of FLASH-RT from the perspective of free radicals, we conducted simulations of reactions occurring during the chemical phase. When ionizing radiation interacts with cellular components or water, it can generate free radicals that subsequently react with biological macromolecules, including lipids, proteins, and DNA, thereby inducing oxidative stress and cellular damage. In this study, protein hydroperoxides (ROOH) was used as a biomarker for radiation-induced damage, as it is not only a product of free radical interactions with biological macromolecules but also a critical mediator in the processes of oxidative stress and cellular damage. In the simulation, MDRs of 900 Gy/s and 0.05 Gy/s were used for FLASH and CONV respectively, consistent with the MDRs used in the experiment. The simulation results (Fig. 3) showed that no radical recombination was observed for CONV, and ROOH increased linearly with dose. For FLASH, the fraction of free radical recombination increased with escalating dose, thus increasing the difference between FLASH and CONV. This means that a certain amount of dose is needed to achieve a given difference in biological effects. Furthermore, the dose delivery time significantly influences the fraction of radical recombination, as illustrated in Fig. 3C. As dose delivery time decreased, the fraction of radical recombination increased, and reached a saturation value when the time was less than 10 ms. The dose delivery time corresponding to the fraction of radical recombination reaching 50% of the maximum value is about 300 ms.

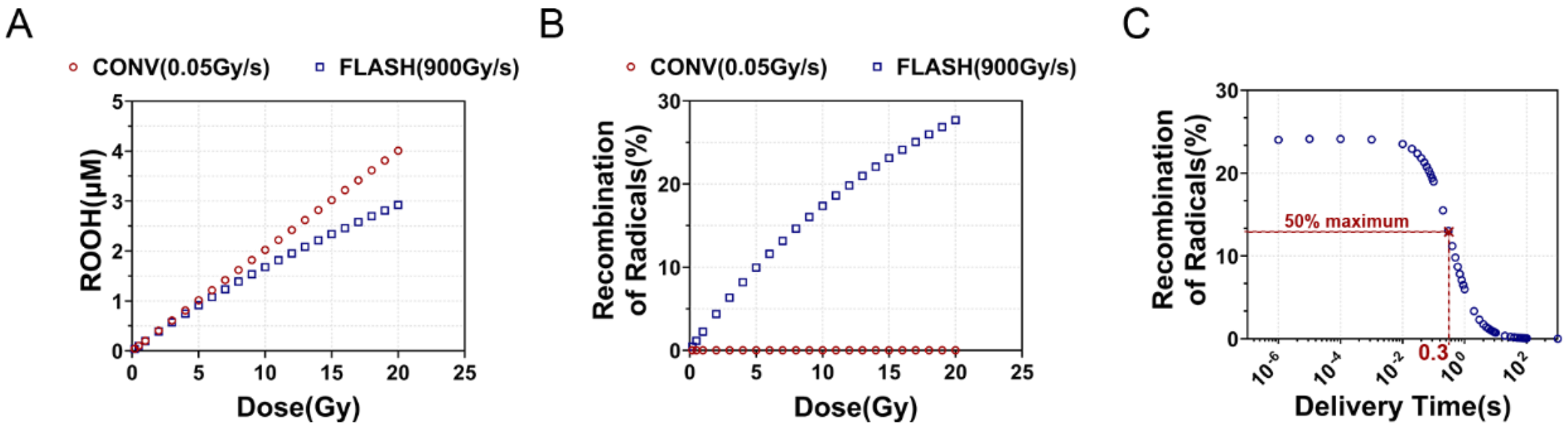

Fig. 3. Simulation results of ROOH concentration and the rates of radical recombination after CONV-RT and FLASH-RT. The concentration of ROOH (A) and the rates of radical recombination (B) were analyzed across various doses, comparing CONV-RT (0.05 Gy/s) with FLASH-RT (900 Gy/s). (C) Rates of free radical recombination vary with the delivery time at a dose of 15 Gy.

## Antioxidants attenuate the disparity in intestinal toxicity observed between the FLASH-RT and CONV-RT groups

The role of free radicals in the sparing effect of FLASH-RT was examined by incorporating antioxidants amifostine. Amifostine, a commonly utilized radioprotective agent, safeguards normal cells from radiation-induced DNA damage via its active metabolite (WR-1065). WR-1065 can eliminate free radicals, thereby reducing the damage caused by radiation. (24). The changes of FLASH group and CONV group after adding antioxidants were compared with MDRs of 750 Gy/s and 0.05 Gy/s, respectively, at a dose of 15 Gy As shown in Fig. 4A, after adding amifostine, it was observed that the structural integrity of the crypts was largely maintained, and the infiltration of inflammatory cells was significantly diminished. After adding amifostine, the radiation damage of CONV group was significantly reduced to a level comparable to that of FLASH group without amifostine, while the change in FLASH group was not significant. These findings indicate that radicals and antioxidants are crucial in mediating the sparing effect observed in FLASH-RT.

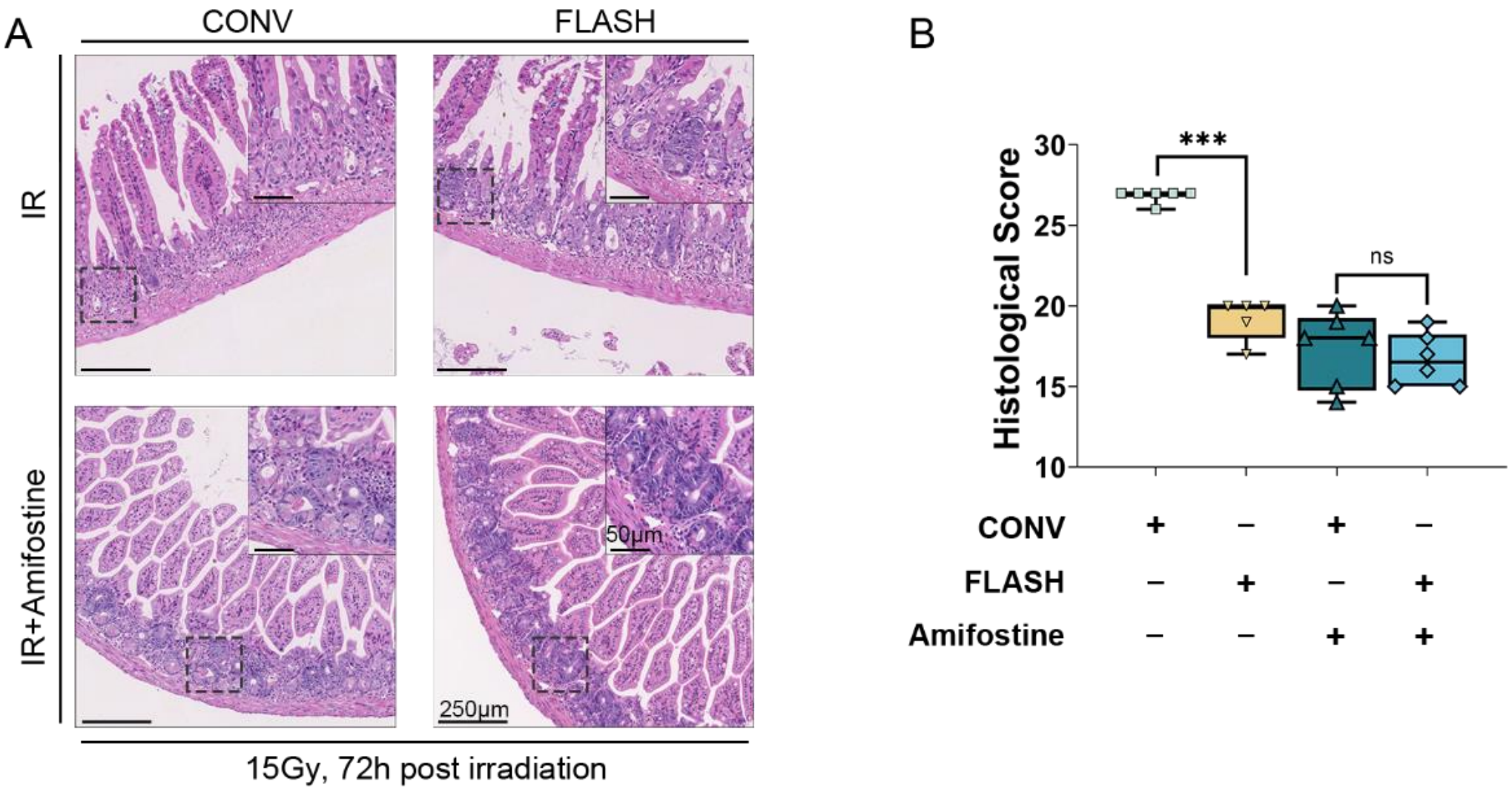

Fig. 4. Comparative analysis of intestinal damage in mice subjected to amifostine administration prior to FLASH-RT or CONV-RT. (A) Representative images of H&E-stained sections of intestinal tissues 72 hours post-treatment, with or without amifostine, at irradiation dose rates of 750 Gy/s and 0.05 Gy/s. (B) Histological scores of mice following irradiation, with or without the administration of amifostine (n=6 per group). Scale Bar = 250 μm. p values are derived from ***p<0.001.

## Discussion

Currently, definitive conclusions regarding the specific physical conditions necessary for observing the FLASH sparing effect at the biological endpoint remain elusive(1, 8, 25). The impact of various physical parameters, including dose, MDR, and IDR, has shown inconsistency across different studies(26–30). Through a quantitative analysis of published experimental data, we determined that both the dose and MDR must surpass certain thresholds to observe the FLASH sparing effect(13). Considering the significant role of free radicals in the damage induced by low linear energy transfer (LET) radiation, we explored the FLASH sparing effect from the perspective of free radicals. Previous studies have proposed the radical recombination hypothesis and the radical recombination-antioxidant hypothesis(20). Nevertheless, there remains a paucity of experimental investigations concerning complex biological entities, such as cells or tissues. In the current study, we examined the critical factors responsible for inducing the FLASH sparing effect, with a particular emphasis on the role of free radicals, utilizing a combination of simulation and experimental methodologies.

As shown in Fig. 3, the simulation results indicated that peroxyl radical recombination did not occur under CONV conditions (0.05 Gy/s), and the concentration of ROOH increased linearly with the dose. For FLASH irradiation with an MDR of 900 Gy/s, the fraction of peroxyl radical recombination increases with the administered dose, thereby amplifying the disparity in ROOH levels when compared to CONV (Fig. 3B). Consequently, the dose must surpass a specific threshold to elicit a discernible difference in biological effects. The experimental findings (Fig. 2B) indicate that, even at an MDR of 900 Gy/s, the FLASH sparing effect was not detected at doses of 6, 9, and 12 Gy, but became apparent at a dose of 15 Gy. Notably, this 15 Gy threshold closely approximates the 50% lethal dose (LD50) of 14.7 Gy associated with acute intestinal toxicity(31). Some studies have also mentioned that the dose threshold of FLASH sparing effect on skin tissue is 25 Gy, which is also close to the LD50 of skin tissue (25-28 Gy)(10, 13, 32). The dose threshold depends on the selected end point because of the varying radiation sensitivity of different end point. Here we chose histological score as a comprehensive indicator in the experiments. In addition, the histological score of FLASH group at the dose of 15 Gy was close to that of CONV group at the dose of 12 Gy. The simulation results indicate that the ROOH levels in the FLASH group at 15 Gy closely approximate those observed in the CONV group at 12 Gy(Fig. 3A), thereby corroborating the hypothesis that peroxyl radical recombination plays a crucial role in the manifestation of the FLASH sparing effect. Furthermore, when maintaining a constant dose and MDR, variations in the DPP and the number of pulses did not yield significant differences in the outcomes (as shown in Fig. 2D), which is consistent with some reported results(26–29). Nevertheless, inconsistent conclusions have been reported regarding the observation of FLASH sparing effect with high DPP and CONV MDR (0.3 Gy/s) using relatively late endpoints(9). These findings are not adequately explained from the perspective of the chemical stage and necessitate further experimental investigation. This also highlights the need for more robust endpoints, such as the LD50.

For a specified dose, the fraction of radical recombination is dependent on the dose delivery time. The simulation results depicting the recombination fraction as a function of dose delivery time at a dose of 15 Gy are presented in Fig. 3C. When the dose delivery time exceeds 10 seconds, the fraction of radical recombination approaches zero. Conversely, when the duration is extremely brief, less than 10 ms, the recombination fraction reaches a saturation level. The dose delivery time at which the recombination fraction achieves 50% of its maximum value is approximately 300 ms. In the dose rate experiment conducted with a dose of 15 Gy (Fig. 2C), the FLASH sparing effect was not detected at a MDR of 40 Gy/s, corresponding to a dose delivery time of 375 ms. However, the effect was observed at an MDR of 100 Gy/s, corresponding to a dose delivery time of 150 ms. In addition, as shown in Fig.2C, the magnitude of the FLASH effect reached saturation after the MDR exceeded 200 Gy/s (i.e., dose delivery time less than 75 ms), which is consistent with the time scale of 100 ms for the saturation of the free radical combination in the simulation.

Upon the addition of antioxidants, a competitive reaction ensues between antioxidants and free radicals, competing with the recombination of free radicals. Experimental findings indicated a significant reduction in acute intestinal injury in the CONV group following the administration of high concentrations of antioxidants, yielding results comparable to the FLASH group devoid of antioxidants. Notably, the FLASH group exhibited minimal variation regardless of antioxidant presence. These observations substantiate the hypothesis that peroxyl radical recombination is the principal factor contributing to the FLASH sparing effect. Furthermore, the levels of antioxidants and antioxidant enzymes are elevated in tumor tissues compared to normal tissues(33), potentially elucidating the similar lethality observed between FLASH and CONV in tumor contexts.

## Conclusions

In the present study, the critical factors responsible for inducing the FLASH effect was elucidated from the perspective of free radicals through a combination of simulation and experimental approaches. Using a single-factor approach, we found that the critical requirement for observing the sparing effect with a specified significance at the biological endpoint is the administration of an adequate dose within the time window of the radical reaction, irrespective of the pulse structure. The relationship between free radical recombination and exposure time and dose were quantitatively analyzed. Additionally, no significant difference in intestinal damage was found between FLASH-RT and CONV-RT after introducing antioxidants, suggesting that the generation and recombination of free radicals are pivotal factors influencing the FLASH sparing effect.

## Conflict of Interest:

None.

## Data Availability Statement

All data generated and analyzed during this study are included in this published article (and its supplementary information files).

## Acknowledgement

Thanks very much for the kind help from Professor Wei Han, who works at Hefei Institutes of Physical Science, Chinese Academy of Sciences. He gave us a lot of advice when performing the in vivo experiment.

## Supplementary Materials

Table. S1 The scoring criteria of the histological score

| | submucosa inflammation | crypts structure | crypts regeneration | inflammation cells infiltration in lamina propria | thickening of the muscularis mucosae | epithelial damage |
|---|---|---|---|---|---|---|
| 0 | no active inflammation and erosions | no significant abnormalities | almost 100% being regenerative crypts | no inflammatory cell infiltration in the lamina propria | no thickening phenomenon | no Paneth cell metaplasia, goblet cell count is normal with no reduction |
| 1 | increased lymphocytes, plasma cells, and eosinophils | crypts are slightly deformed, with a reduction in numbers | 80-90% are regenerative crypts | a slight increase in lymphocytes and plasma cells | slight thickening | slight or localized Paneth cell metaplasia |
| 2 | a significant neutrophil infiltration | some crypts show deformation | the number of regenerative crypts begins to decrease significantly. | moderate infiltration of inflammatory cells in the lamina propria, such as neutrophils and eosinophils. | moderate thickening, with obvious fibrosis and muscularization. | Paneth cell metaplasia is evident, affecting multiple areas |
| 3 | the degree of infiltration further increases | crypt structure is markedly altered | 80% of crypts are deformed and have lost regenerative capacity. | severe infiltration of inflammatory cells in the lamina propria | further thickening | Paneth cell metaplasia is widely distributed. |
| 4 | extensive mucosal damage and | Crypts are completely deformed | Almost no regenerative crypts within the | Rare neutrophil infiltration of crypt epithelial cells, with | Severe thickening. | Massive Paneth cell metaplasia occurs, with a reduction of |

| | | | | | |
|---|---|---|---|---|---|
| inflammatory cell infiltration | and atrophic. | section range. | further deepening of the degree of inflammatory cell infiltration. | | goblet cells at the tips of the villi, and the epithelial layer may be accompanied by extensive erosion or ulceration. |